\documentclass[apjl]{emulateapj}

\usepackage{graphicx}
\usepackage{apjfonts}

\slugcomment{Accepted in ApJ, in press}
\shorttitle{First detection of apparent swirling motions in the feet of solar prominences}
\shortauthors{Orozco Su\'arez, Asensio Ramos, and Trujillo Bueno}

\newcommand{\degree}{\ensuremath{^\circ}\/}
\newcommand{\kms}{~km~s$^{-1}$}

\begin{document}

\title{EVIDENCE FOR ROTATIONAL MOTIONS IN THE FEET OF A QUIESCENT SOLAR PROMINENCE}

\author{D.\ Orozco Su\'arez\altaffilmark{1,2}, A.\ Asensio Ramos\altaffilmark{1,2}, and J.\ Trujillo Bueno\altaffilmark{1,2,3}}

\email{dorozco@iac.es}

\altaffiltext{1}{Instituto de Astrof\'isica de Canarias, E-38205 La Laguna, Tenerife, Spain}
\altaffiltext{2}{Departamento de Astrof\'isica, Universidad de La Laguna, E-38206 La Laguna, Tenerife, Spain}
\altaffiltext{3}{Consejo Superior de Investigaciones Cient\'ificas, Spain}

\begin{abstract}
We present observational evidence of apparent plasma rotational motions in the feet of a solar prominence. Our study is based on spectroscopic observations taken in the \ion{He}{1}~1083.0~nm multiplet with the Tenerife Infrared Polarimeter attached to the German Vacuum Tower Telescope. We recorded a time sequence of spectra with 34 s cadence placing the slit of the spectrograph almost parallel to the solar limb and crossing two feet of an intermediate size, quiescent {\it hedgerow} prominence. The data show opposite Doppler shifts, $\pm$~6\kms, at the edges of the prominence feet. We argue that these shifts may be interpreted as prominence plasma rotating counterclockwise around the vertical axis to the solar surface as viewed from above. The evolution of the prominence seen in EUV images taken with the {\it Solar Dynamic Observatory} provided us clues to interpret the results as swirling motions. Moreover, time-distance images taken far from the central wavelength show plasma structures moving parallel to the solar limb with velocities of about $10-15$\kms. Finally, the shapes of the observed intensity profiles suggest the presence of, at least, two components at some locations at the edges of the prominence feet. One of them is typically Doppler shifted (up to $\sim$~20\kms) with respect to the other, thus suggesting the existence of supersonic counter-streaming flows along the line-of-sight.   
\end{abstract}

\keywords{Sun: chromosphere --- Sun: filaments, prominences}

\section{Introduction}
\label{Section1}

Prominences are one of the most conspicuous solar structures embedded in the chromosphere or in the hot solar corona (for reviews see \citealt{2010SSRv..151..243L} and \citealt{2010SSRv..151..333M}). They are made of cool magnetized plasma suspended above the solar surface. Recently, the understanding of prominence dynamics has greatly benefited from the high cadence and continuity of data taken with instruments such as the extreme ultraviolet light telescope (AIA; \citealt{2012SoPh..275...17L}) onboard NASA's {\it Solar Dynamics Observatory} ({\it SDO}; \citealt{2012SoPh..275....3P}). For instance, AIA observations show that the feet of  solar prominences seem to be swirling in the lower corona \citep{2012arXiv1205.3819L}. Unfortunately, it is not possible to infer plasma velocities from AIA narrow band images. The only possibility is to measure local displacements of intensity structures in the plane of sky from consecutive images. This technique gives a measure of the plasma velocity in the plane perpendicular to the line-of-sight (LOS), provided the measured displacements can be considered as real plasma motions. It was recently used by \cite{2010ApJ...716.1288B} to determine transverse velocities of dark bubbles in quiescent filaments. Other authors have measured LOS plasma velocities from ground-based spectroscopic observations using the H$_{\alpha}$ line  (e.g., \citealt{2009ApJ...704..870L,2005SoPh..226..239L,1998Natur.396..440Z,1988A&A...197..281S,1983A&A...119..197M,1981SoPh...69..301M}) or lines in the EUV range (e.g., \citealt{1985SoPh...96...35E,1979SoPh...61...39V,1976ApJ...210L.111L}). 

\begin{figure}
\begin{center}
\epsscale{1}
\plotone{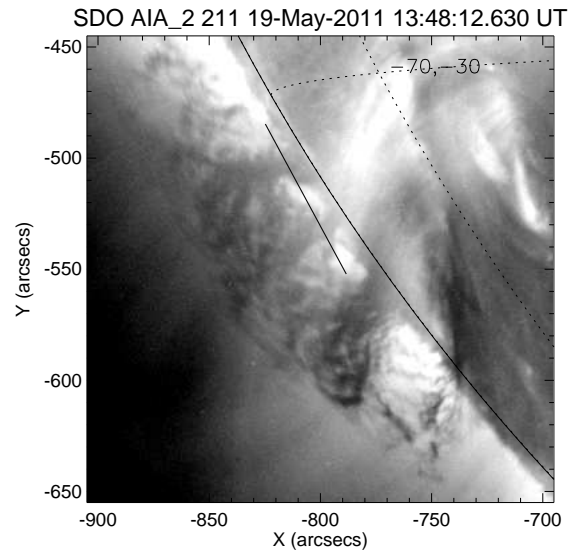}
\end{center}
\caption{{\it SDO}/AIA \ion{Fe}{14} 211 \AA\/ image of the quiescent {\it hedgerow} prominence. The straight line represents the TIP-II spectrograph slit, inclined $8.8$\degree\/ with respect the solar limb and at about 10\arcsec\/ from it. The online journal mpeg animation shows the evolution of one of the prominence feet in the \ion{Fe}{9}~171~\AA\/ band-pass during the TIP-II observations.}
\label{fig1}
\end{figure}

\begin{figure}
  \centering
\epsscale{1}
\plotone{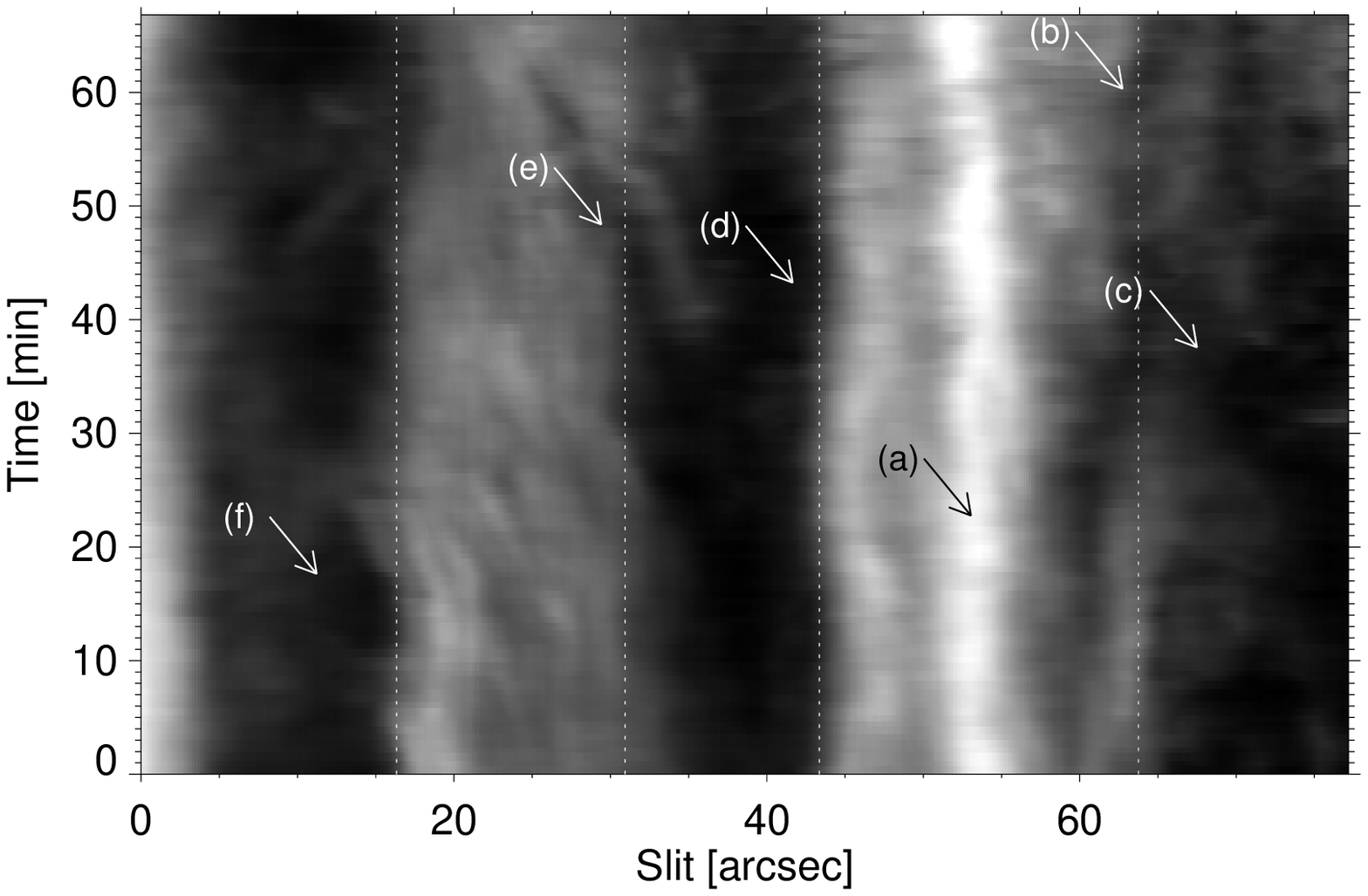}  
\plotone{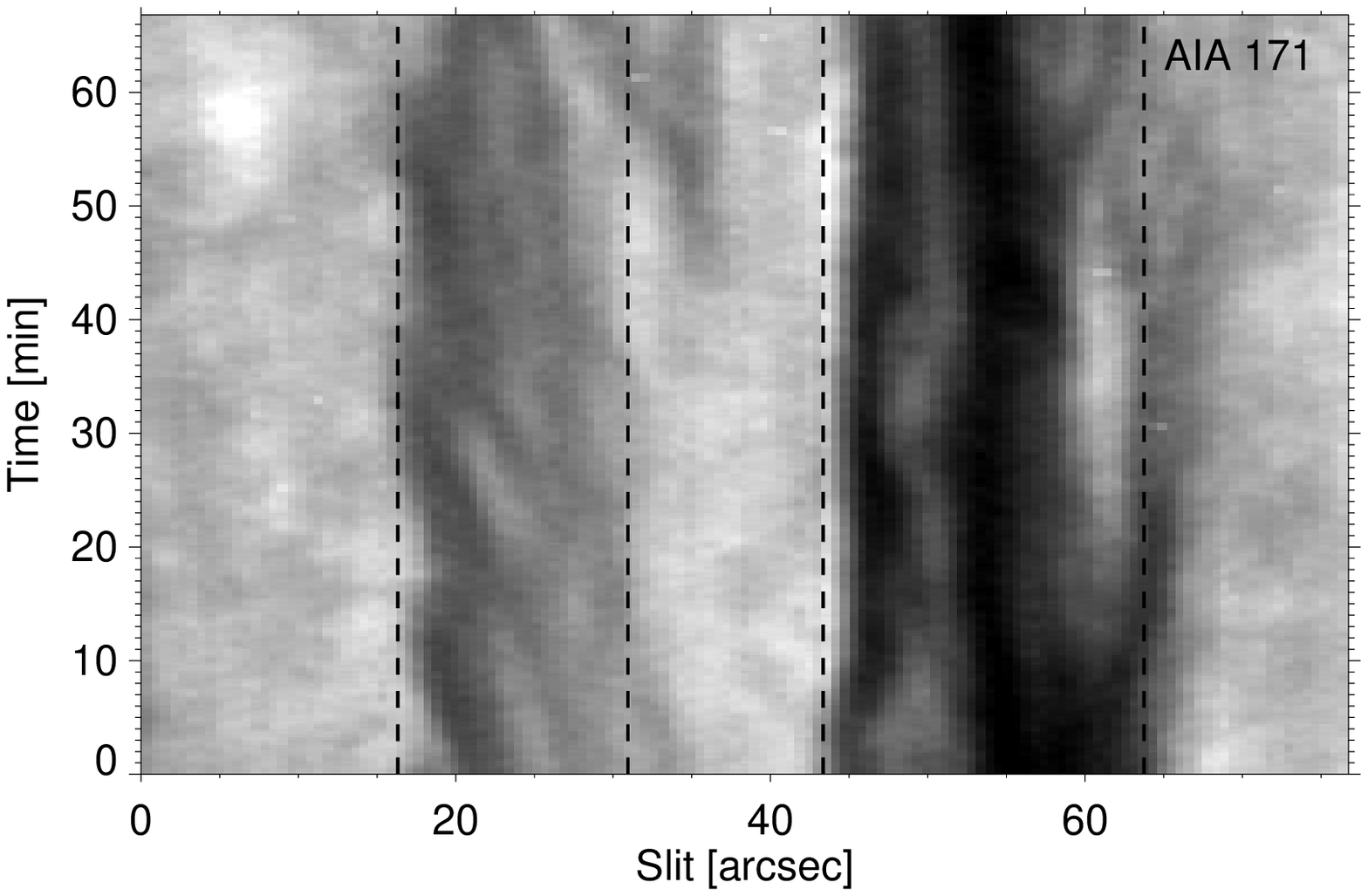}
\caption{Top panel: temporal variation (Y-axis) of the \ion{He}{1} 1083.0~nm line peak intensity along the slit (X-axis). The feet of the prominence are the bright structures against a dark background. Bottom panel: time-slice diagram generated with the 171~\AA\/ pass-band AIA data at the same location of the TIP-II slit. The feet of the prominence are seen as dark structures against a bright background in the AIA data. The resemblance between both maps suggests that AIA and TIP-II data were aligned successfully.}
\label{fig2}
\end{figure}

In this Letter, we measure spectral line Doppler shifts in the feet of an off-limb quiescent solar prominence using ground-based spectropolarimetric observations in the \ion{He}{1}~1083.0~nm triplet. We first present the observations and then determine Doppler shifts from time-averaged and individual profiles. We also determine plasma shifts parallel to the slit. We argue that these observations show evidence of swirling motions in the feet of the observed prominence.

\section{Observations}
\label{sec2}

The observations were taken between 12:53 UT and 13:58 UT on 2011 May 19 with the Tenerife Infrared Polarimeter (TIP-II; \citealt{2007ASPC..368..611C}) installed at the German Vacuum Tower Telescope (VTT) of the Observatorio del Teide (Tenerife, Spain). TIP-II measured the four Stokes parameters of the \ion{He}{1}~1083.0~nm multiplet in sit-and-stare mode. In particular, we placed the spectrograph slit (0\farcs5 wide and 80\arcsec\/ long with 0\farcs17 sampling) nearly parallel to the solar limb (forming an angle of $8.8$\degree\/ with it). The exposure time per slit position and polarization state was 7.5 s. The final cadence was 34 s and the total duration of the time sequence was 66.3 minutes (corresponding to 117 slit images). The 1083.0~nm spectral region sampled by the spectrograph slit spanned 0.726~nm in 1.1~pm bins, thus including the \ion{Si}{1}~1082.7088~nm and an atmospheric water vapor line at 1083.21~nm. 

The TIP-II data reduction process includes dark current, flat-field, fringes correction, and polarimetric calibration. The absolute wavelength calibration was done using the atmospheric H$_2$O line of the averaged spectrum as a reference. Then, we corrected the solar and terrestrial rotation and the relative shifts between the Sun and Earth orbits. This correction varies 0.36~pm during the observations, introducing a systematic error in the wavelength of $\pm$100~m~s$^{-1}$ in 1 hr. Overall, the absolute error in the Doppler shifts is $\sim$325~m~s$^{-1}$.

In our observations, the TIP-II slit crossed two feet of a quiescent {\it hedgerow} prominence located at the solar south east (Figure~\ref{fig1} sketches the position of the slit on a {\it SDO}/AIA \ion{Fe}{14} 211 \AA\/ snapshot). The position of the slit was kept rather stable during the observations thanks to the high performance of the Kiepenheuer-Institute Adaptive Optics System installed in the VTT \citep{2003SPIE.4853..187V}. The rms of the displacements of the slits in the plane of the sky should be less than 1 arcsecond. The evolution of one of the prominence feet can be seen in an \ion{Fe}{9} 171 \AA\/ AIA movie\footnote{In the movie, arrows represent the TIP-II slit.}, available in the electronic edition of this paper in the {\it Astrophysical Journal}. The animation shows the apparent rotational motion of the  prominence feet.

The temporal variation of the \ion{He}{1} 1083.0~nm line peak intensity is displayed in the top panel of Figure~\ref{fig2}. Note first that the \ion{He}{1}~1083.0~nm triplet is seen in emission in off-limb prominences. The slit time sequence samples two feet of the prominence, the one on the right side is much more intense than the one on the left side. The data were obtained with relatively good seeing conditions. Indeed, the feet show a glimpse of the fine-scale in the prominence. We estimate a spatial resolution of about 1\arcsec\/ using cross-correlation techniques.

The bottom panel of Figure~\ref{fig2} displays a time-slice diagram generated with the {\it SDO}/AIA \ion{Fe}{9} 171 \AA\/ data at the same location of the TIP-II slit. The position of the slit with respect to the AIA data was estimated using the orientation of the slit with respect the plane of the zero solar meridian, the angle between the slit and the line tangent to the nearest limb position, and H$_\alpha$ slit-jaw images. The co-alignment error should be within the AIA resolution limit of 1\farcs5. In the AIA data the feet of the prominence are seen in absorption mainly from neutral hydrogen and neutral and ionized helium against a bright background. The \ion{He}{1} 1083.0~nm peak intensity and the AIA absorption show similar spatial and temporal variations. The reason for the good resemblance between the EUV and infrared data is that the EUV spectrum below 504~\AA\/ ionizes helium which recombines again producing an enhancement of the populations in its triplet states and in particular in the metastable level (2s$^3$S$_1$) of the transition. This produces an increase in the emission of the \ion{He}{1} 1083.0~nm multiplet, which would be hardly explainable with collisional excitation alone \citep{1997ApJ...489..375A,2008ApJ...677..742C}.

We used STEREO-B EUVI 195~\AA\/ observations \citep{2008SSRv..136....5K} to estimate the filament orientation (prominences are seen in absorption when seen on disk). The prominence spine was rotated  $\sim17$\degree\/ counterclockwise with respect to the solar south-north direction. Stereo data also helped to determine the solar longitude of the prominence, about 80\degree\/ East. 

\section{Calculation of Doppler shifts in the prominence feet}
\label{sec3}

\subsection{Average  Doppler shifts}

\begin{figure}
  \centering
\epsscale{1}
\plotone{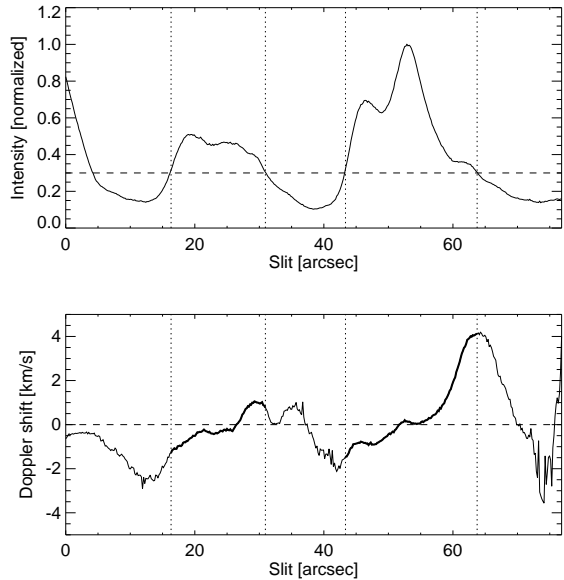}  
\caption{Top panel: variation of the time-averaged peak intensity of the \ion{He}{1}~1083.0~nm line along the slit, normalized to its maximum value. Bottom panel: Doppler shifts inferred from the \ion{He}{1} triplet. In both panels, the vertical lines delimit the feet of the prominence (see text).}
\label{fig3}
\end{figure}

We first restrict the analysis to the temporally averaged intensity profiles. Figure~\ref{fig3} shows the spatial variation of the peak intensity of the temporally averaged spectra along the slit. The intensity variation exhibits two large bumps corresponding to the prominence feet (one from 15\arcsec\/ to 30\arcsec\/ and another from 45\arcsec\/ to 64\arcsec). Note that the peak intensity of the \ion{He}{1}~1083.0~nm triplet never reaches zero in these data. Hence, we constrained our analysis to those pixels whose peak intensity surpasses 30\% of the maximum signal (dashed line in top panel from Figure~\ref{fig3}). The Doppler shifts are derived from the intensity profile by fitting a Gaussian. We fit three parameters: the Doppler shift, the amplitude of the Gaussian, and its width. 

The bottom panel of Figure~\ref{fig3} shows the inferred Doppler shifts,  where positive means redshifts. If we focus on the selected pixels only, we readily notice that the left sides of the feet are blue-shifted whereas the right sides are red-shifted. The inferred Doppler shift smoothly varies from one edge to the other edge, being at rest at about the center of the feet. The red-shifts are greater in the right feet, around slit positions $\sim$60\arcsec. There, the \ion{He}{1}~1083.0~nm peak intensity decreases rapidly and the peak signal falls below the 30\% threshold at 64\arcsec, coinciding with a dramatic decrease in the Doppler shift.

\subsection{Doppler analysis of individual profiles}
\label{3.2}

\begin{figure*}[!t]
  \centering
\epsscale{0.38}
\plotone{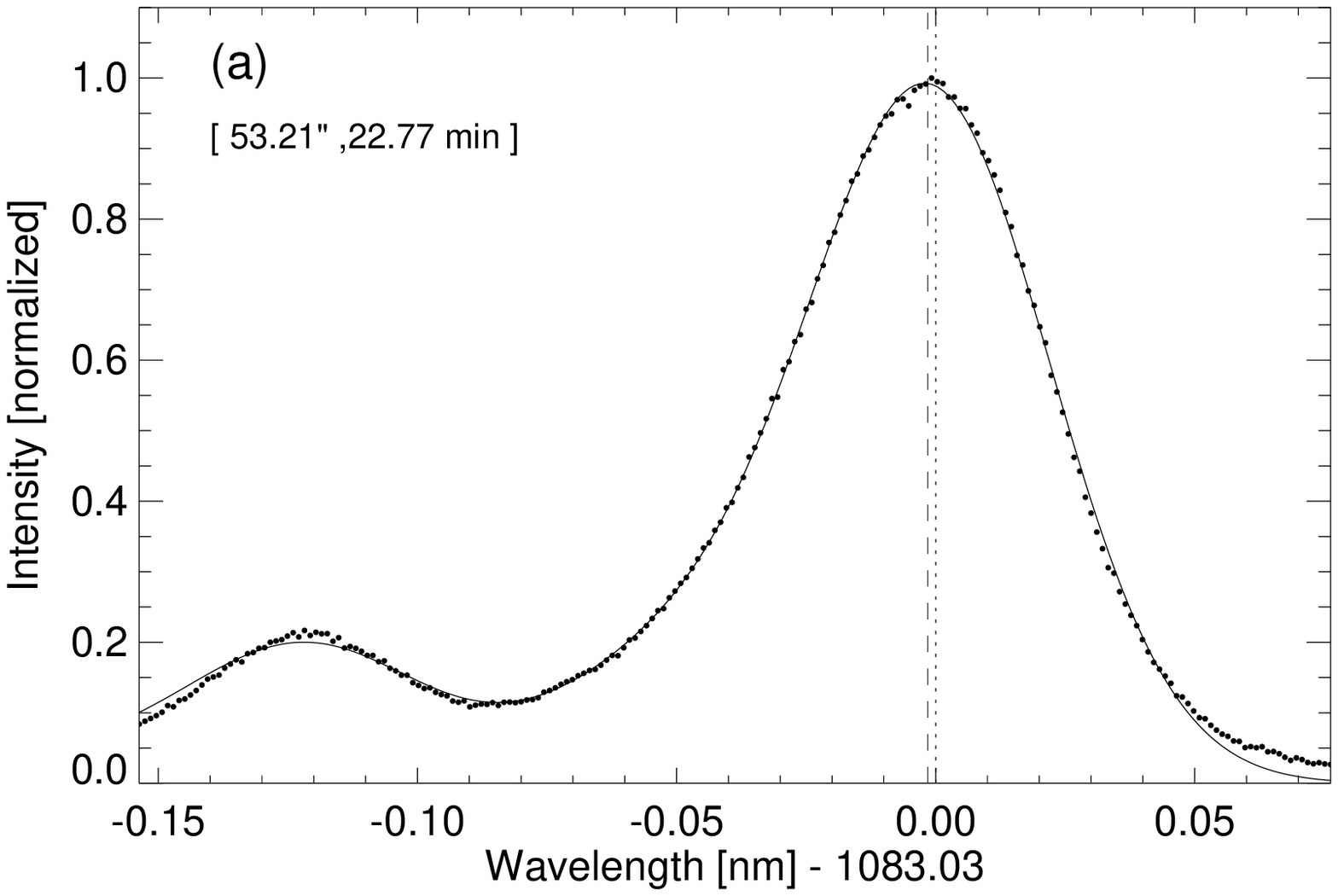}  
\plotone{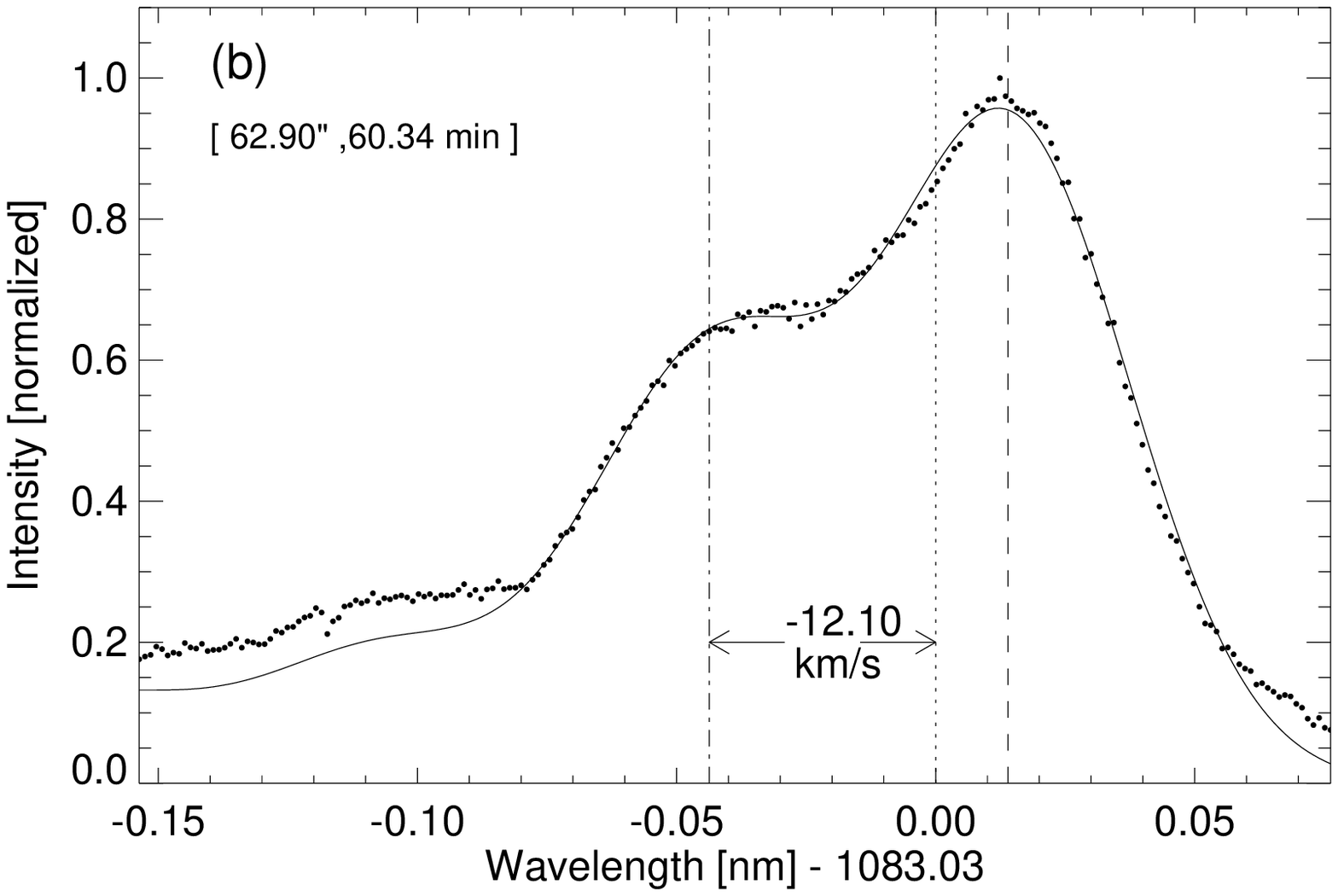}  
\plotone{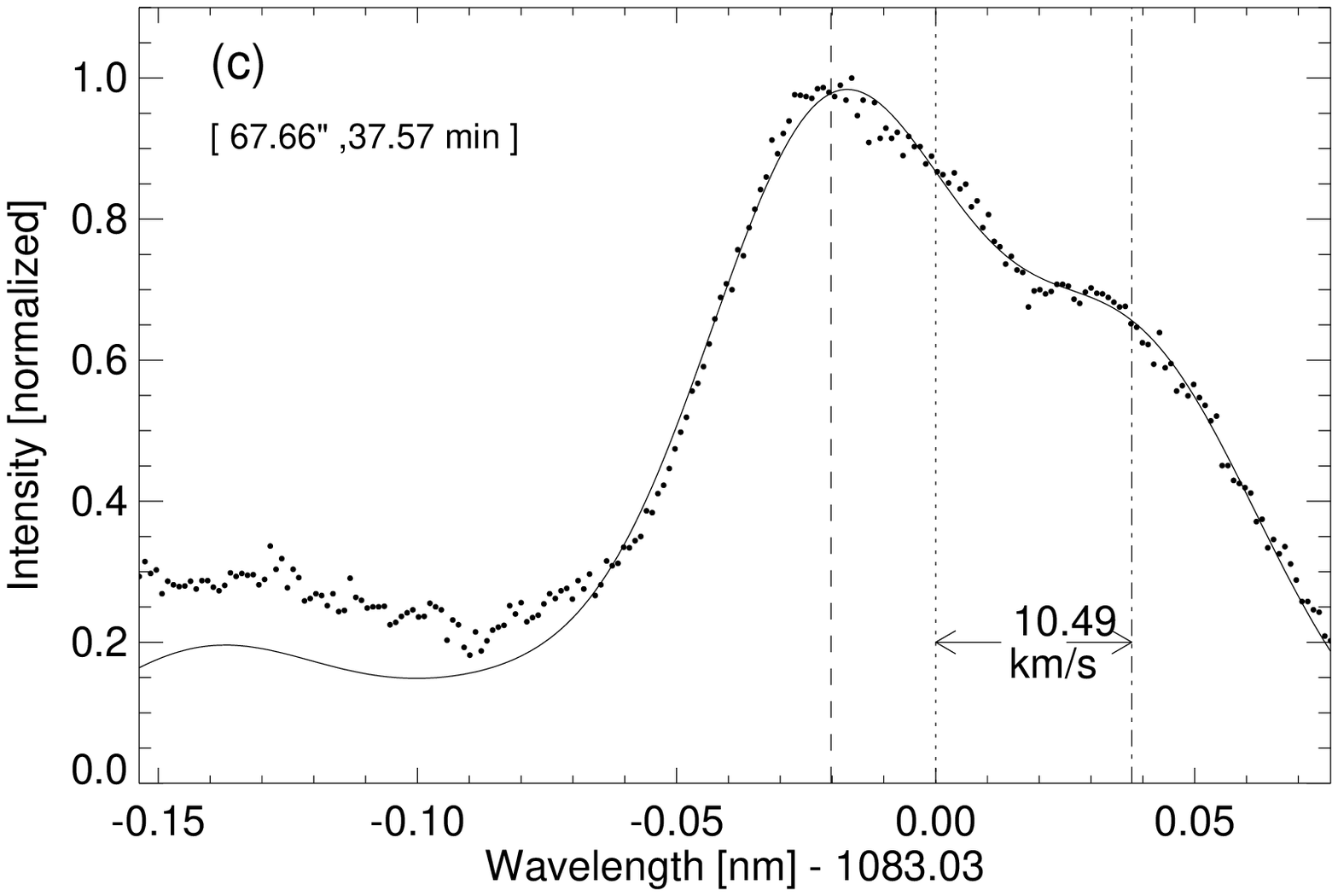}  
\plotone{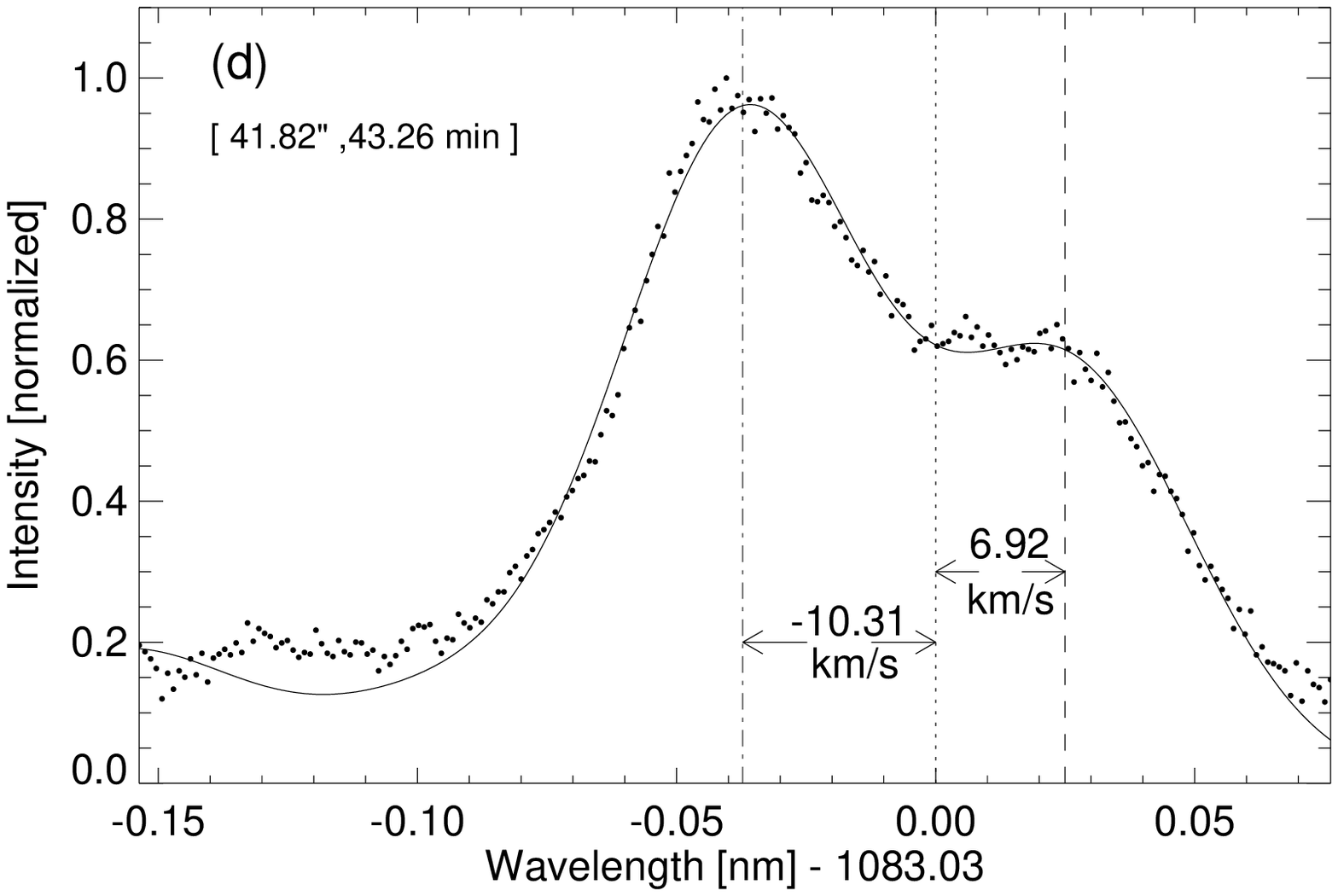}  
\plotone{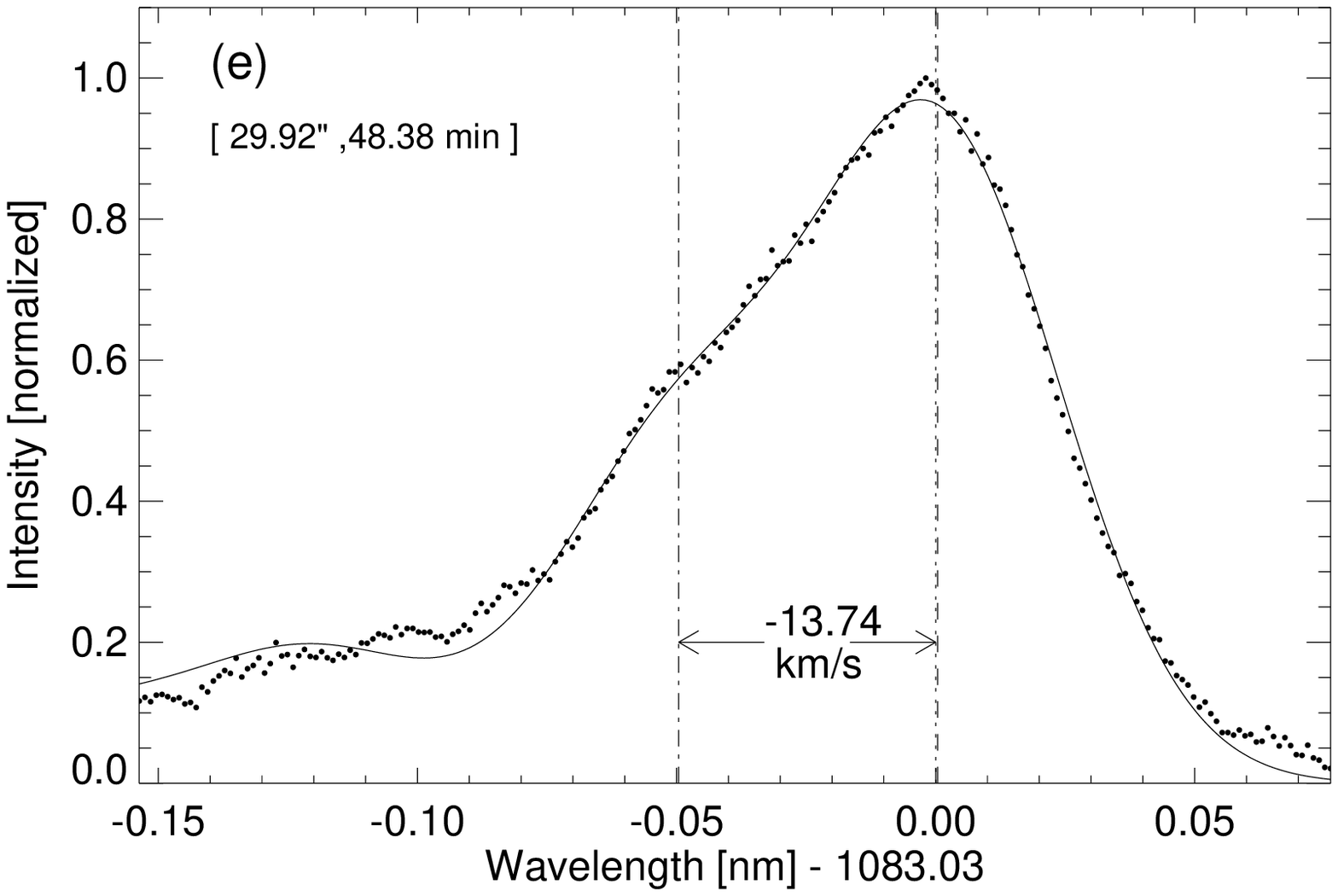}  
\plotone{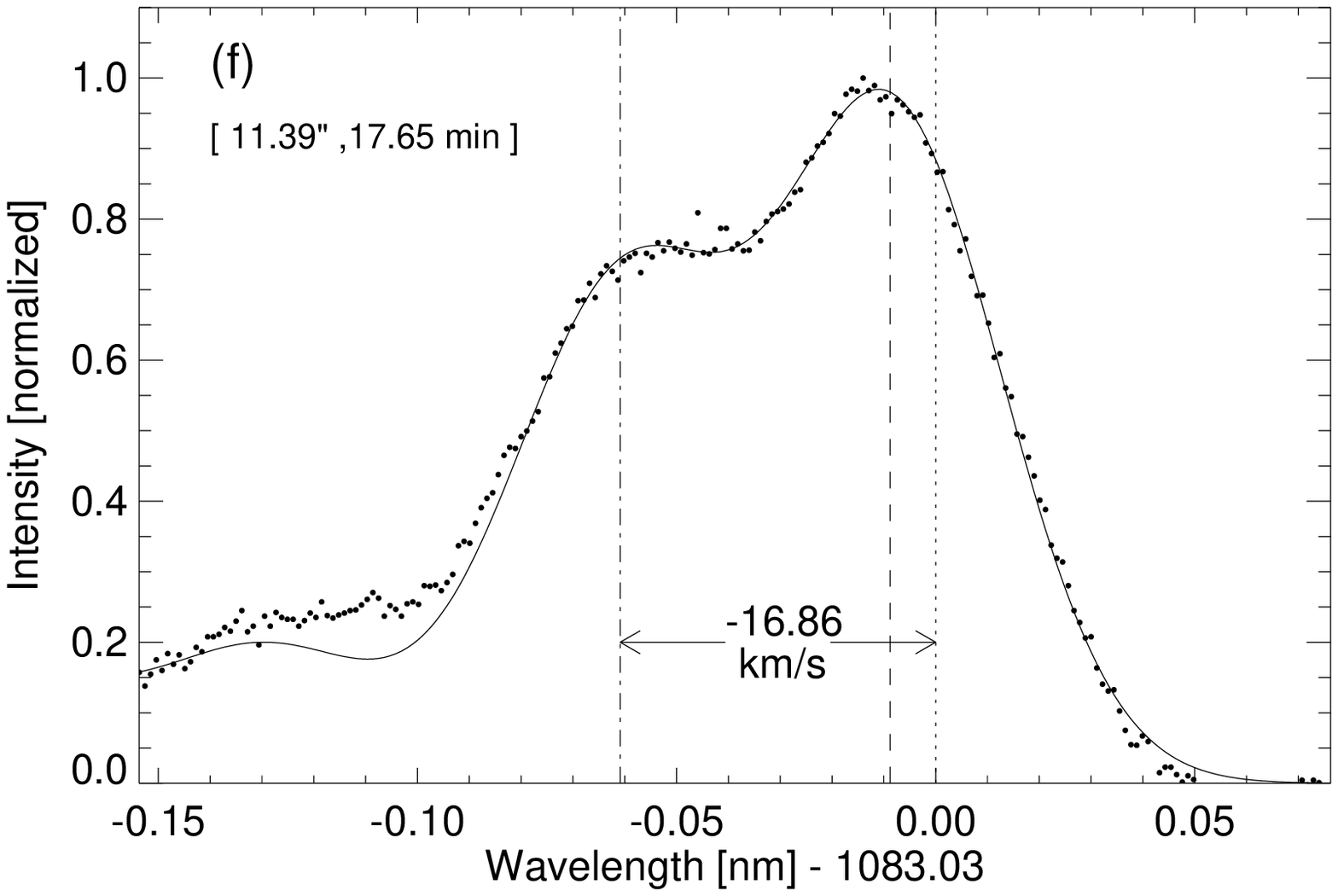}  
\caption{Example \ion{He}{1} 1083.0~nm individual profiles taken from different regions of the prominence. Panel (a) represents a prototypical example of a rest profile, showing the fine structure of the multiplet, with a weak blue component at 1082.909~nm ($^3$S$_1-^3$P$_0$) separated about 0.12~nm from the other two components at about 1083.029~nm ($^3$S$_1-^3$P$_1$ and $^3$S$_1-^3$P$_2$), which are indistinguishable. Panels (b), (c), (d), (e), and (f) display individual profiles showing, at least, two-components. Dots represent the observations and solid a two-component Gaussian fit. Arrows in Figures~\ref{fig2}~and~\ref{fig4} show their  location with respect the prominence feet and the corresponding observing time. The dot-dashed and dashed vertical lines stand for the wavelength location of the shifted and main components. The dotted lines show the rest position.}
\label{fig4}
\end{figure*}

\begin{figure*}[!t]
  \centering
\epsscale{1}
\plotone{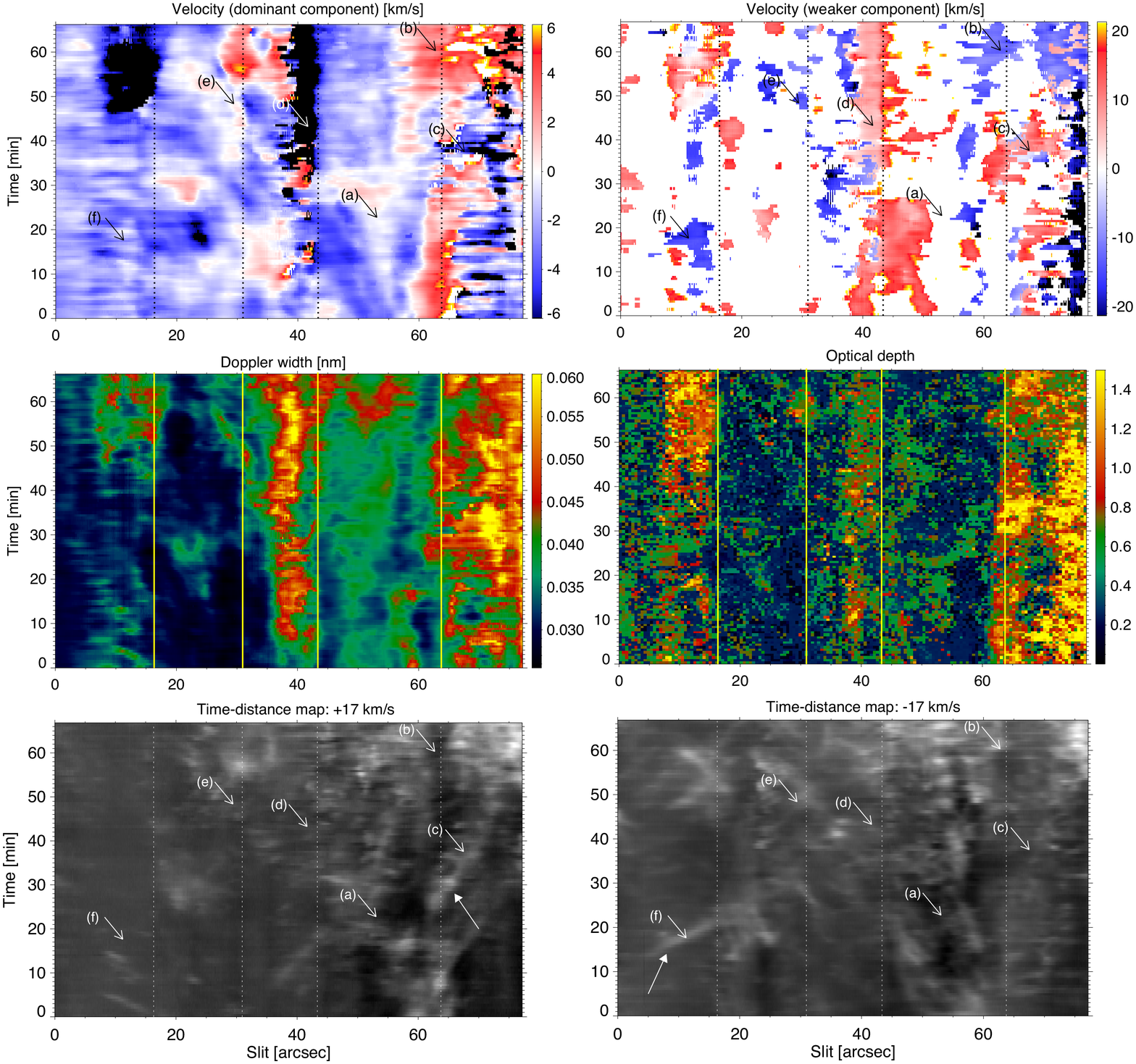}  
\caption{Top panels: Doppler shifts resulting from the two-components Gaussian fit of the intensity profiles. Left panel represents the shifts of the dominant component, saturated at $\pm$6\kms, whereas the right panel shows that corresponding to the weaker component, saturated at $\pm$20\kms\/ (see Sect.\ref{3.2}). White areas in the top-right panel represent locations where the profiles show one dominant component only. Left middle panel displays the Doppler width of the dominant component retrieved from the two Gaussian component fit. Right middle panel represents the optical depth of the \ion{He}{1}~1083.0~nm multiplet inferred from a one component inversion with HAZEL. Bottom panels: time-distance maps taken at $\pm$~17\kms\/ (left and right, respectively for + and -) from the \ion{He}{1} line center (1083.029~nm). Vertical lines delimit normalized intensity values larger than 0.3 units (see Figure~\ref{fig3}).  Arrows show the location of profiles displayed in Figure~\ref{fig4}. Arrows with solid arrowheads show locations of transverse Doppler shift patterns (Section~3.3).}
\label{fig5}
\end{figure*}

The visual inspection of the individual emission profiles corresponding to the prominence feet revealed that they show, at least, two different components. The one showing the greatest peak intensity (hereafter dominant component) is generally slightly blueshifted or redshifted depending on the location of the pixel with respect to the center of the prominence foot. The other (hereafter weaker) component showing a lesser peak intensity is conspicuously Doppler shifted. Examples of these profiles are displayed in Figure~\ref{fig4}. Their location is marked in Figures~\ref{fig2}~and~\ref{fig5}. Profile (a) represents a prototypical, one-component \ion{He}{1}~1083.0~nm emission profile, i.e., showing two peaks corresponding to the red and blue components of the \ion{He}{1} fine-structure. Profile (b), located in the right side of the right foot, shows two components. In this case, the dominant component moves away from the observer in agreement with the Doppler pattern derived from the temporally averaged profiles while the other component is, surprisingly, strongly blue-shifted; i.e., moving towards the observer at about $12$\kms. Not all profiles in that part of the prominence show the same behavior but there are few exceptions in isolated places. For instance, the dominant component in profile (c) shows an opposite Doppler shift to that inferred from the temporally averaged profiles. Profile (d) also shows two components and strong Doppler shifts. In this case the profile belongs to the left side of the right foot; i.e., where the plasma moves toward the observer. Here, the dominant component is blue-shifted while the weaker component is red-shifted. In this example the Doppler shifts of the dominant component are quite relevant, up to $\sim10$\kms. Profile (e), belonging to the left foot, shows a similar behavior as profile (b) although here the shift of the dominant component is less conspicuous, probably because the intensities in the left foot are overall weaker than the right foot, while the weaker component shows a strong Doppler shift, of $\sim -14$\kms. Finally, the dominant component of profile (f), corresponding to the left side of the left feet, shows Doppler shifts opposite to that inferred from the temporally averaged profiles.

The Doppler shifts have been determined by a two-components Gaussian fit to the observed spectral lines. The fits can be seen superposed on the profiles shown in Figure~\ref{fig4}. The fits are good although sometimes the blue component of the \ion{He}{1} triplet is miss-fitted, probably because of the simplicity of the model approach. To model the fine-structure of the \ion{He}{1} triplet we assumed that each of the two components consisted of three individual Gaussians with the same Doppler width, one for the blue component of the line and two for the red component of the line. For simplicity, we fix the ratio between the blue and the red components of the triplet to 5. However, this ratio may change depending on the amount of ionizing EUV coronal irradiation and on the line opacity \citep{2008ApJ...677..742C}, hence producing the misfit shown in the example profiles of Figure~\ref{fig5}. To select pixels with two plasma components we set a threshold in the full width at half maximum of the observed line profiles because profiles with two components tend to be much wider than the rest. This threshold is completely arbitrary but direct inspection of the data shows that most two-component profiles were selected. 

The inferred Doppler shifts from the two-component Gaussian fit are displayed in Figure~\ref{fig5}. The Doppler shift pattern of the dominant component (top-left panel) shows that the two sides of the feet move in opposite direction. This behavior persists in time, being much stronger at the end of the observations (see top of upper left panel in Figure~\ref{fig5}) where the Doppler shift reaches $\pm$6\kms. The Doppler shift pattern matches the one obtained from the temporally averaged profiles. The Doppler shift map corresponding to the weaker component (top-right panel) shows red-shifts on the left side of the feet and blue-shifts when they are found on the right side of the feet, although exceptions exist, e.g., areas around pixels (c) and (f). The Doppler shifts may be as large as 20\kms. Interestingly, the Doppler shifts exhibited by these weaker components are of opposite sign to that obtained from the dominant component. 

In Figure~\ref{fig5} we also display the Doppler width of the dominant component inferred from the two-component Gaussian fit. The Doppler widths are about a factor of 1.5 larger than that expected from thermal broadening only, about 0.023 nm at 10\,000K. Thus the observed emission profiles may be broadened because of the presence of unresolved components, either along the LOS or within the pixel resolution. Here we distinguish, at least, two components. Figure~\ref{fig5} also shows the optical thickness of the \ion{He}{1}~1083.0~nm emission line. It has been estimated from a one component fit to the data using the HAZEL inversion code \citep{2008ApJ...683..542A}. They are of the order of 0.2$\sim$0.6 around the position of the prominence feet. This means that the \ion{He}{1}~1083.0~nm line is optically thin and therefore we cannot discard that the Doppler shifts we are measuring might be associated with real plasma velocities along the LOS.

\subsection{Transverse shifts}

Figure~\ref{fig5} (bottom panels) represent {\it time-distance} maps taken in the blue and red side around the \ion{He}{1} line center (1083.029~nm) and corresponding to Doppler shifts of $\pm$~17\kms. To enhance the temporal variation of the structures we have removed the averaged intensity profiles along the slit in the two maps. Both maps show intensity patterns moving along the slit. These patterns,  often associated with flows, allow us to infer transverse shifts, i.e., plasma displacements parallel to the solar limb and perpendicular to the LOS. Two clear examples of such patterns are pinpointed by arrows with solid arrowheads in Figure~\ref{fig5}. The slope of the structures implies horizontal speeds of $\sim$10\kms\/ and 15\kms. Interestingly, the profiles associated with these structures, profiles (e) and (f), are those where the dominant component showed a Doppler shift opposite to that obtained from the analysis of the temporally averaged spectra.

\section{Discussion}
\label{sec4}

The Doppler shifts inferred from the temporally averaged intensity profiles show that the left and right sides of the prominence feet {\it move} in opposite directions, with a smooth variation from $- 2$\kms\/ at the left edge to $4$\kms\/ at the right edge in the right foot and from $- 1$\kms\/ at the left edge to $1$\kms\/ at the right edge in the left foot.
We also inferred Doppler shifts from individual profiles using a two-components Gaussian fit. The results for the dominant component show speeds of about $\pm6$\kms\/ at the edges of the prominence feet, again with a smooth variation along the slit. The sign of the Doppler shifts corresponding to the dominant component coincides with that obtained from the temporally averaged data. The inferred Doppler pattern is similar in both prominence feet and the pattern persisted during the whole observation in the right feet. 

To determine whether the measured Doppler shifts can be ascribed to real plasma velocities we inferred the optical thickness of the observed spectral lines and their Doppler widths. The optical thickness reveals that the observed spectral lines are optically thin while the inferred Doppler widths suggest that there might be unresolved components within the spatial resolution limit and/or along the LOS. In these data two of these components are clearly distinguishable in some areas at the prominence feet. Hence, we believe that the inferred Doppler shifts are not contaminated by the superimposition of fine structure elements. Therefore, we assume that the observed \ion{He}{1} 1083.0~nm triplet is tracing real plasma velocities. The fact that the time evolution and spatial pattern showed by the EUV and \ion{He}{1} 1083.0~nm data are similar strengthens this assumption. 

Since in the EUV movie one sees plasma apparently swirling around the prominence feet axis we argue that the inferred Doppler shifts are consistent with the presence of prominence plasma structures rotating counterclockwise around the vertical axis to the solar surface as seen from above. The inferred Doppler shifts along the slit would correspond to the LOS component of such rotational motion. The rotation is maintained during the whole observation. If the prominence feet are cylindrical, vertical structures with an average radius of 20\arcsec, it would take about 4 hr to complete a rotation at an angular speed of 6\kms. This rotation speeds coincide with those inferred in so-called solar tornados using indirect methods \citep{2012arXiv1205.3819L,2012Natur.486..505W,2012arXiv1208.0138S}. We also detect plasma shifts parallel to the slit using time-slice diagrams.

Finally, the results show the presence of strong LOS flows (associated with the weaker component of the observed profiles) localized in areas surrounding the prominence feet. The Doppler shifts reach up to $\pm20$\kms\/ and have sign opposite to the Doppler shifts determined from the dominant component of the profiles. Interestingly, these flows are close to supersonic. The sound speed in prominence plasma taking into account partial ionization and electron densities in the range from $10^{9}$ to $10^{11}$ cm$^{-3}$ and temperatures between 6000 and 10\,000~K ranges from 8.5\kms\/ to 14\kms\/ (c.f., \citealt{2010ApJ...716.1288B}). The inferred Doppler shifts represent lower limits since they correspond to the LOS component only. We have no clues about the origin of such strong LOS flows, although, the most probable scenario is the existence of supersonic counter-streaming flows \citep{1998Natur.396..440Z} along the line-of-sight in the prominence feet. The physical origin of the counter-streaming flows may be the Rayleigh-Taylor instability recently found in prominences \citep{2008ApJ...676L..89B}. In this case, we would be detecting a combination of hot (upward) and cold (downward) plasma flows resulting from the Rayleigh-Taylor instability. Since these instabilities have been found acting in the direction perpendicular to the solar surface, projection effects may be also playing a role.

\acknowledgments
We thank  M.\ Collados and M.\ J.\ Mart\'inez Gonz\'alez for their invaluable help in the TIP-II data reduction and B.\ Ruiz Cobo and A.\ D\'iaz for constructive criticisms of the manuscript. Financial support by the Spanish Ministry of Economy and Competitiveness (MINECO) through the project AYA2010-18029 (Solar Magnetism and Astrophysical Spectropolarimetry) is gratefully acknowledged. AAR also acknowledges financial support through the Ram\'on y Cajal fellowship. Based on observations in the VTT operated on the island of Tenerife by the KIS in the Spanish Observatorio
del Teide of the Instituto de Astrof\'isica de Canarias.

\end{document}